\documentclass[a4paper]{jpconf}
\usepackage{graphicx}
\usepackage{orcidlink}
\usepackage{subcaption}
	\hypersetup{ 
	colorlinks=false,
	linkcolor=black,
	anchorcolor=black,
	citecolor=black,
	filecolor=black,
	menucolor=black,
	runcolor=black,
	urlcolor=black,
	hidelinks=true
}
\begin{document}

\title{\href{https://indico.cern.ch/event/855454/contributions/4596351/}{Signal to background discrimination for the production of double Higgs boson events via vector boson fusion mechanism in the decay channel with four charged leptons and two b-jets in the final state at the LHC experiment}}

\author{\href{https://orcid.org/0000-0002-9361-3142}{Brunella D'Anzi}$^{1,2}$, \href{https://orcid.org/0000-0002-0625-6811}{Nicola De Filippis}$^{2,3}$, \href{https://orcid.org/0000-0001-7069-0252}{Walaa Elmetenawee}$^{1,2}$ and \href{https://orcid.org/0000-0002-4018-0128}{Giorgia Miniello}$^{1,2}$}

\address{$^1$ Department of Physics, University of Bari Aldo Moro, Via E. Orabona n.4, I-70126 Bari,Italy }
\address{$^2$ Istituto Nazionale di Fisica Nucleare (INFN), Via E. Orabona n.4, I-70126 Bari, Italy}
\address{$^3$ Politecnico di Bari, Via Amendola, 126/B, I-70126 Bari, Italy }

\ead{\href{mailto:brunella.danzi@ba.infn.it}{brunella.danzi@ba.infn.it}}

\begin{abstract}
At the CERN Large Hadron Collider experiment, the non-resonant double Higgs production via vector-boson fusion represents a unique mean to probe the VVHH (V=Z, W$^\pm$) Higgs self-coupling at the current center of mass energies. Such a rare signal cannot be separated efficiently from huge backgrounds by applying a few-observables cut-based selection. Indeed, in this work, a Deep Learning algorithm is used to decide whether an event is more signal- or background-like. In particular, we report on two main aspects: results on a hyper-parameters parallel scanning strategy to distribute the training process across multiple nodes on the ReCaS-Bari data center computing resources and on the discriminating performance of a Deep Neural Network architecture.
\end{abstract}

\section{Introduction and Related Work}

Nowadays High Energy Physics (HEP) analyses take advantage of Multivariate Analysis (MVA) techniques to optimize the discrimination between signal and background, preserving as much signal as possible. Indeed, running a classical cut-based selection on several physics observables would imply a severe reduction of both signal and background candidates, which would turn out to be a quite inefficient choice especially when signal events are rare as usually happens when performing Higgs-related studies.
On the other hand, HEP MVA focuses on using a pre-determined set of independent variables optimally combined to build discriminants which could effectively separate signal from background. 

In this context, an Artificial Neural Network naturally implements a MVA since it is a complex Deep Learning computing system that mimics the biological neural networks receiving these variables as input, training and testing on them and, eventually, producing a single output for binary classification problems \cite{Peterson:1991wf,Glorot:2010}. 
Generally, a Neural Network consists of nodes organized in layers. Each node receives either a feature of the problem or a weighted sum of the previous layer node output. It uses several hyper-parameters (e.g. learning rate, number of hidden layers, neurons per each layer, dropout fraction, the minimizer algorithm of the loss function, etc.) that are usually set manually in order to define the network architecture. 

In our work, we used Monte Carlo generated events from non-resonant Higgs boson pair production analysis at the energies of the LHC, where one of the Higgs bosons decays into the four-lepton final state and the other one decays into a pair of b quarks as described in \autoref{signaltopology}. The signal and background data sets were generated to run the analysis on data corresponding to the integrated luminosity reached by the Large Hadron Collider (LHC) experiment at CERN during the 2018 Run II period using proton-proton collisions at a center of mass energy of 13 TeV. Furthermore, a DNN classifier, usually considered a very versatile and efficient method for these kind of problems, has been implemented by using the open-source library Keras and the open source platform for machine learning TensorFlow \cite{tensorflow2015-whitepaper,Keras}. 
In particular, we show in \autoref{comments} the implementation and optimization of a DNN classifier for discrimination of our signal from SM background events. We then conclude with a brief outlook on our plans for future development.
\section{The Signal Topology}\label{signaltopology}
The discovery of the Higgs boson at the Large Hadron Collider in 2012 \cite{CMS:2012qbp,ATLAS:2012yve} opened a new frontier in HEP both for Standard Model (SM) and Beyond Standard Model (BSM) scenarios. 
\begin{figure}
	\begin{subfigure}{0.2\textwidth}
		\includegraphics[width=\linewidth]{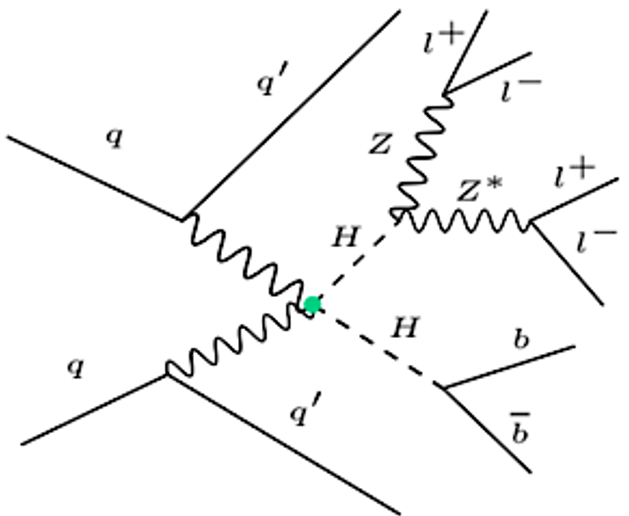}
		\caption{} \label{fig:2a}
	\end{subfigure}\hspace*{\fill}   
	\begin{subfigure}{0.2\textwidth}
		\includegraphics[width=\linewidth]{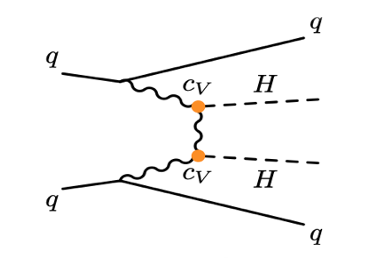}
		\caption{} \label{fig:2b}
	\end{subfigure}\hspace*{\fill}   
	\begin{subfigure}{0.2\textwidth}
		\includegraphics[width=\linewidth]{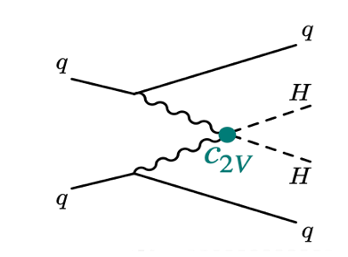}
		\caption{} \label{fig:2c}
	\end{subfigure}\hspace*{\fill}   
	\begin{subfigure}{0.2\textwidth}
		\includegraphics[width=\linewidth]{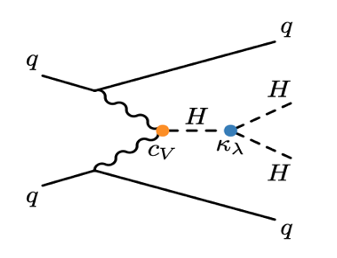}
		\caption{} \label{fig:2d}
	\end{subfigure}\hspace*{\fill}   
	\caption{\small{(\ref{fig:2a}) Feynman diagram of the leading order HH vector boson fusion signal physics process for this data analysis; (\ref{fig:2b}), (\ref{fig:2c}), (\ref{fig:2d}) show the different double Higgs production mechanisms which are included in the Standard Model and Beyond Standard Model theories with different values for the Higgs self-couplings C$_{2V}$, C$_{V}$, $\kappa_{\lambda}$. }}\label{fig:2}
\end{figure}
Thus, a new era of ambitious studies has been opened up. After the precise measurement of the main parameters of the SM Higgs, one of the most important objectives to be accomplished is the measurement of the Higgs self-couplings, which are strictly related to the shape of the Higgs potential (see \autoref{fig:2}).

The Higgs self-coupling studies clearly involve the investigation of the pair production of Higgs bosons \cite{potentialatcolliders}. In contrast to what happens for the dominating production mechanism of gluon-gluon fusion, the production of two Higgs bosons via a vector-boson fusion (VBF) mode turns out to be a particularly important process for the determination of the quartic-Higgs coupling  C$_{2V}$. In fact, in the VBF mechanism, the Higgs bosons are produced at leading order from heavy gauge bosons that are radiated off two quarks, which can be used as tags for jets to simplify the experimental identification and measurement.
Thus, the innovation for this study is related both to the technological point of view and to the originality of the physics analysis signature considered. In fact, differently from the single Higgs boson production modes, widely explored and studied during the Run I and II at the LHC, the double Higgs boson production via VBF in the four lepton plus 2 b-jets final state has not yet been investigated. This was mainly due to the small value of its cross section weighted with the branching ratios. Indeed, for the HH production via VBF, with the Higgs mass set to its best fit value of 125.09 GeV, the cross section at 13 TeV is around 1.723 fb \cite{crossectionsfirst} while the corresponding branching ratios are 2.79 $\times$ 10$^{-4}$ for H $\to$ ZZ$^{*}$ $\to$ 4l, with l = e, $\mu$, $\tau$, and 5.75 $\times$ 10$^{-1}$ for H $\to$ $b\overline{b}$ \cite{crossections}. This requires an exclusive event selection in order to efficiently perform a background rejection mostly from SM single Higgs processes and HH gluon-gluon fusion events. Nevertheless, the analysed double Higgs boson decay mode is highly interesting because of the four charged lepton Higgs boson decay mode and the two b-jets final state inclusion. The  former is one of the highest signal-to-background (S/B) ratio and has an excellent invariant mass resolution, while the latter has the highest branching ratio among the Higgs boson decay modes. 

\begin{center}
\begin{figure}
	\begin{subfigure}{0.35\textwidth}
	\includegraphics[width=\linewidth]{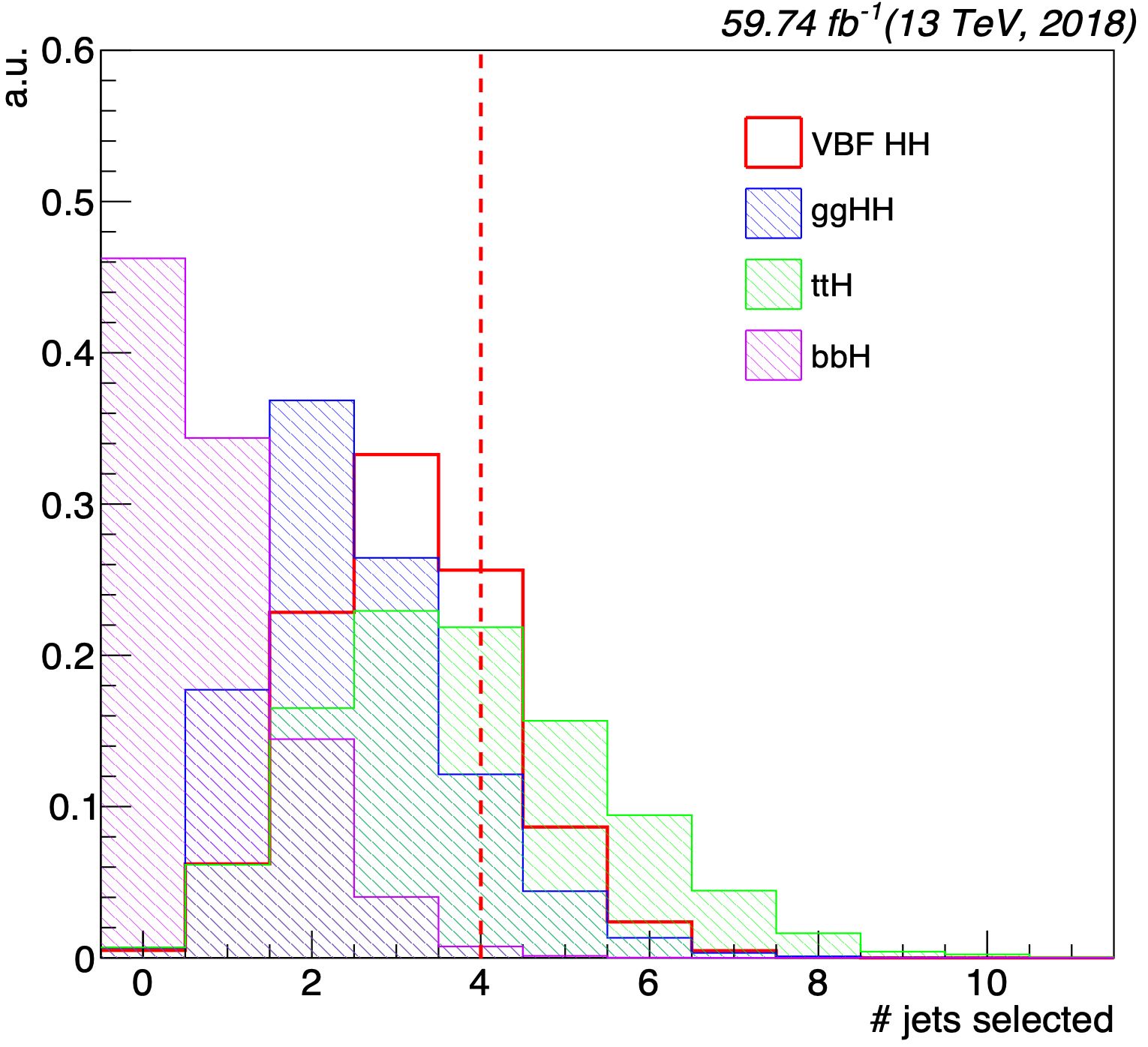}
	\caption{} \label{fig:3a}
	\end{subfigure}\hspace*{\fill}   
	\begin{subfigure}{0.38\textwidth}
	\includegraphics[width=\linewidth]{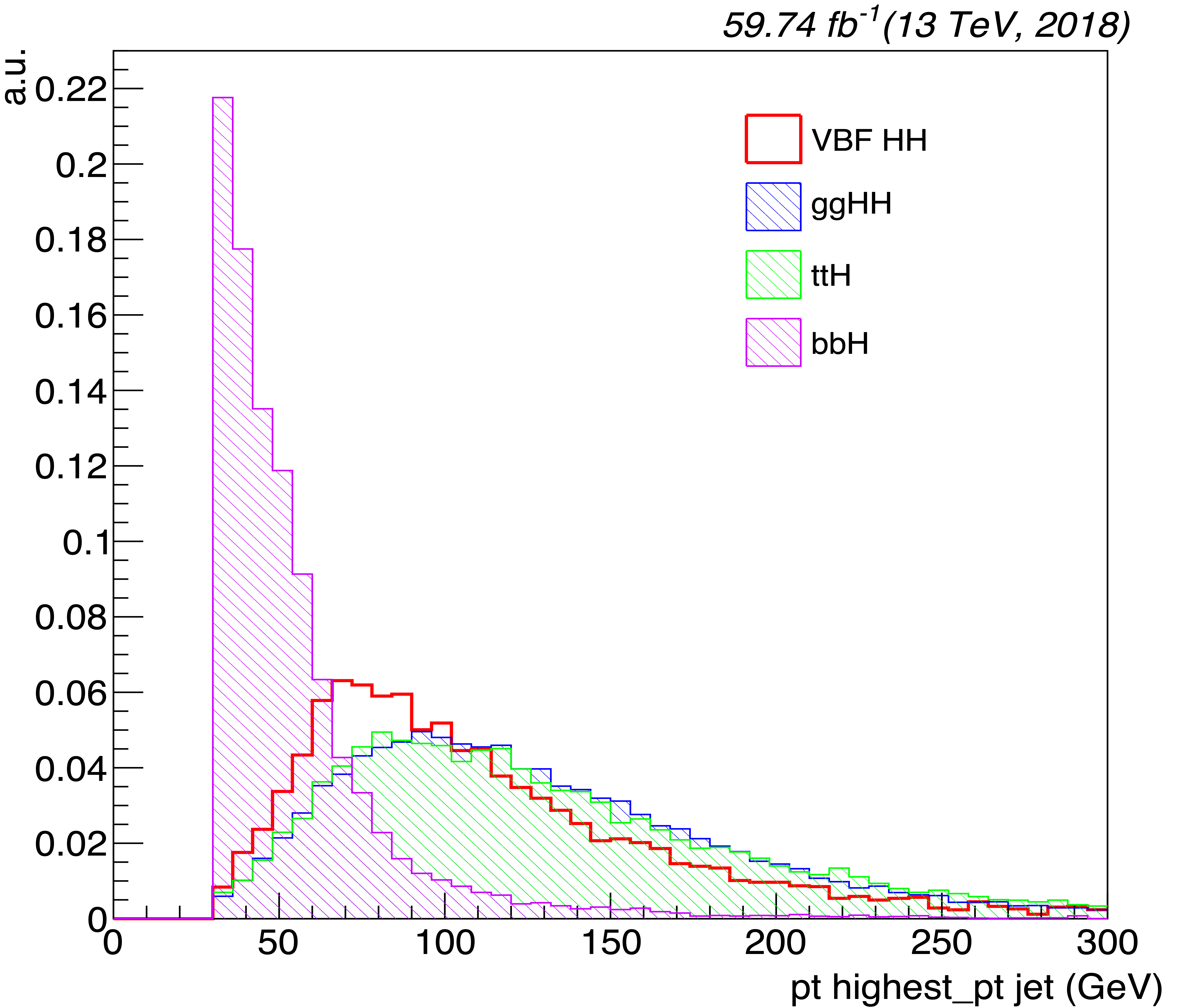}
	\caption{} \label{fig:3b}
	\end{subfigure}\hspace*{\fill}\\ 
	\begin{subfigure}{0.37\textwidth}
	\includegraphics[width=\linewidth]{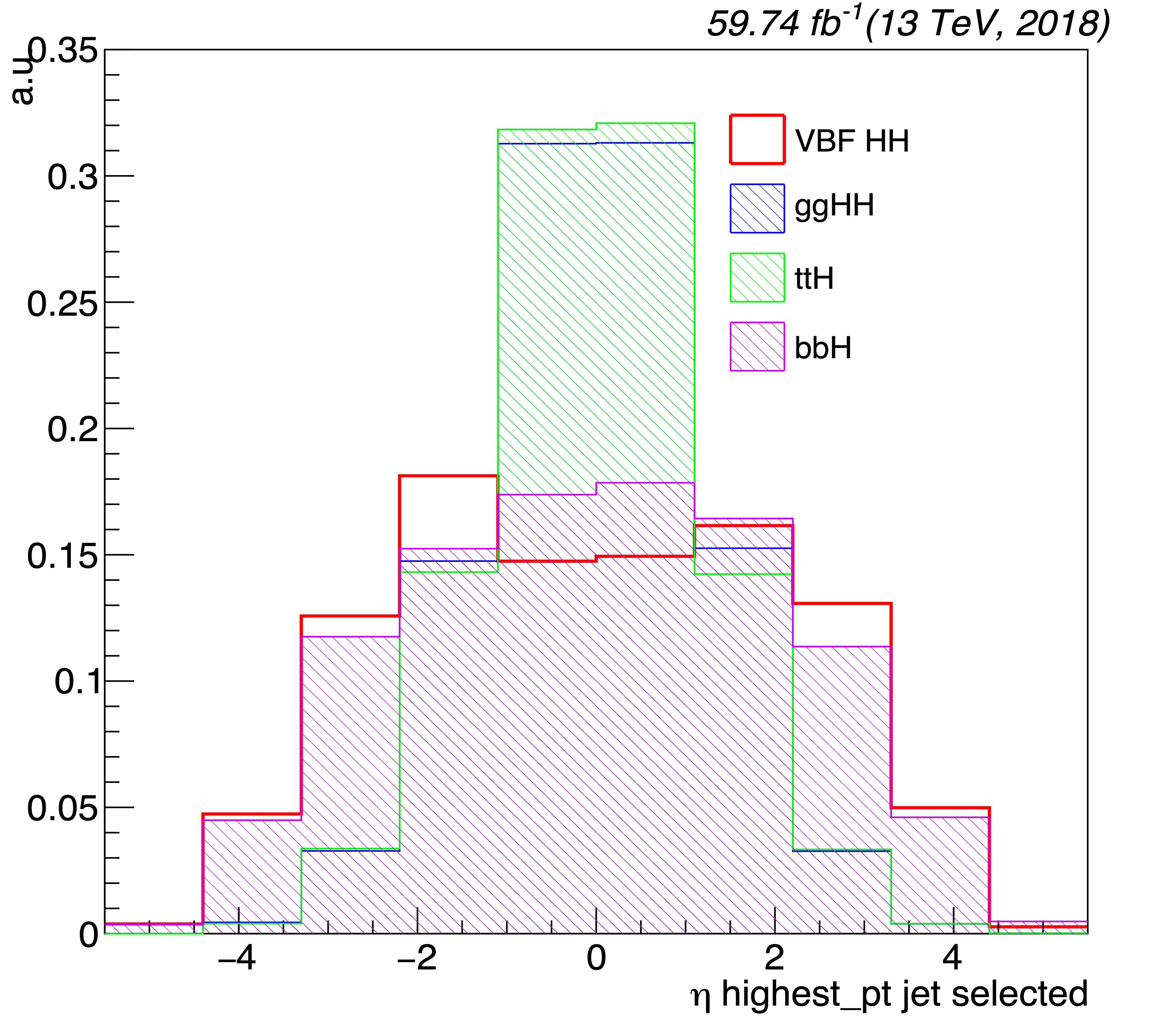}
	\caption{} \label{fig:3c}
	\end{subfigure}\hspace*{\fill}   
	\begin{subfigure}{0.37\textwidth}
	\includegraphics[width=\linewidth]{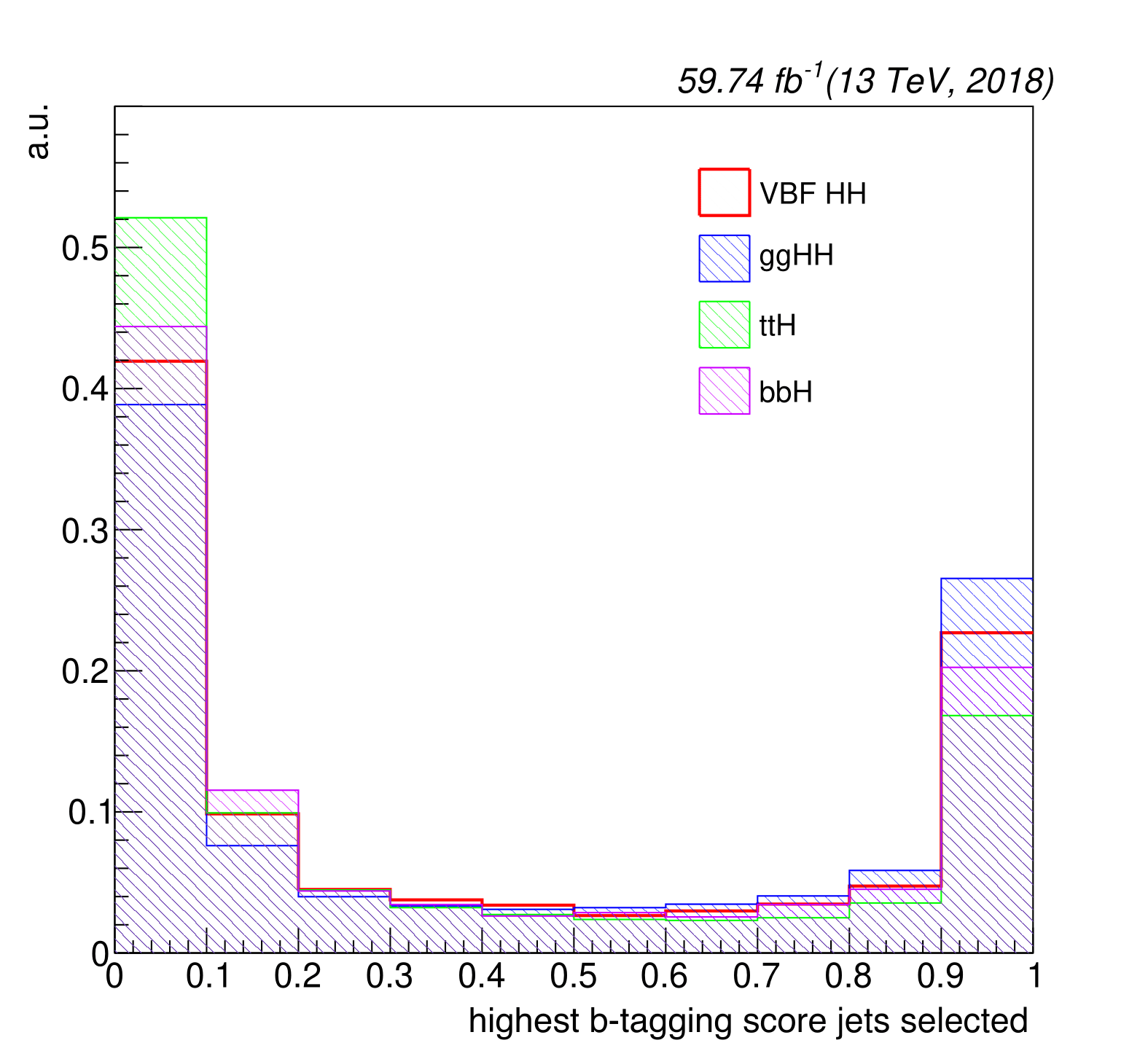}
	\caption{} \label{fig:3d}
	\end{subfigure}\hspace*{\fill}   
	\caption{\small{Examples of normalized signal and main backgrounds data set distributions for some physical observables: (\ref{fig:3a}) shows the number of jets passing the requirements in \autoref{signaltopology}; the DNN supervised learning tool is trained by having as inputs the kinematic observables p$_{T}$, $\eta$, $\phi$ for the four charged leptons and for the six highest transverse momentum $p_{T}$ (\ref{fig:3b}, \ref{fig:3c}) jets plus their b-tagging score (\ref{fig:3d}). }} \label{fig:3}
\end{figure}
	\end{center}

In particular, the analysis has been applied to signal and background events for the channels 4e, 4$\mu$ and 2e2$\mu$ separately and the merged samples are used for training the Deep Learning algorithm. To properly prepare the data sets, we defined a signal region corresponding  to a portion of the phase space which is expected to be populated with signal events from the model of interest, the VBF double Higgs production, while having low background rates. The event passes the signal selection if it includes at least one primary vertex, a Z candidate having an invariant mass 12 $< m_{ll(\gamma)} < $  120 GeV/c$^{2}$, a ZZ candidate from a pair of Z bosons, which do not have common leptons (non-overlapping), a number of jets higher than 4 (see \autoref{fig:3}), $\eta <$4.7 and an angular separation between each jet and lepton of $\Delta R_{jl} >$ 0.3.

\begin{figure}
\begin{center}
	\hspace*{\fill}\begin{subfigure}{0.49\textwidth}
		\includegraphics[width=\linewidth]{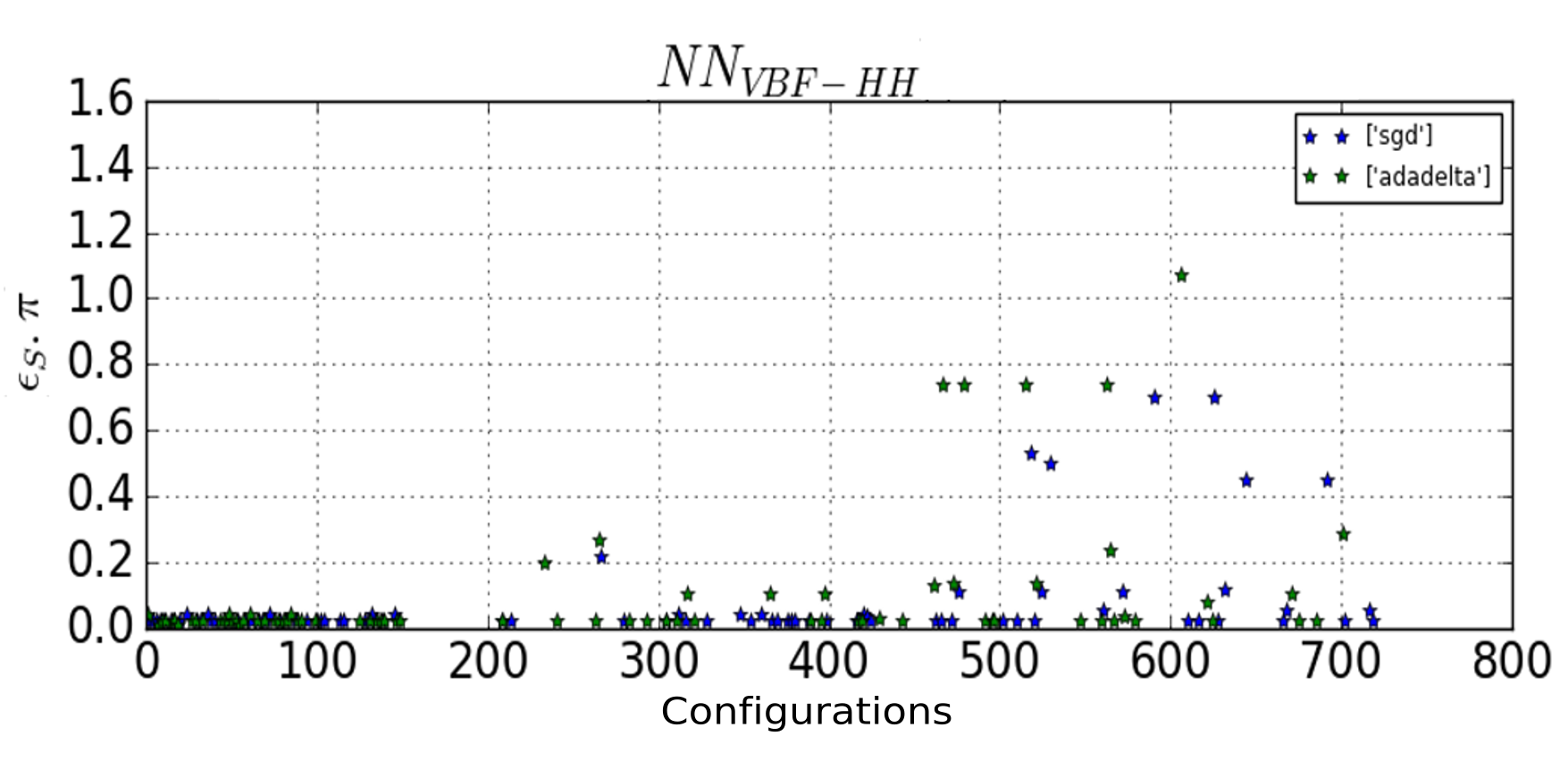}
		\caption{} \label{fig:5a}
	\end{subfigure}\hspace*{\fill}  
	\begin{subfigure}{0.49\textwidth}
		\includegraphics[width=\linewidth]{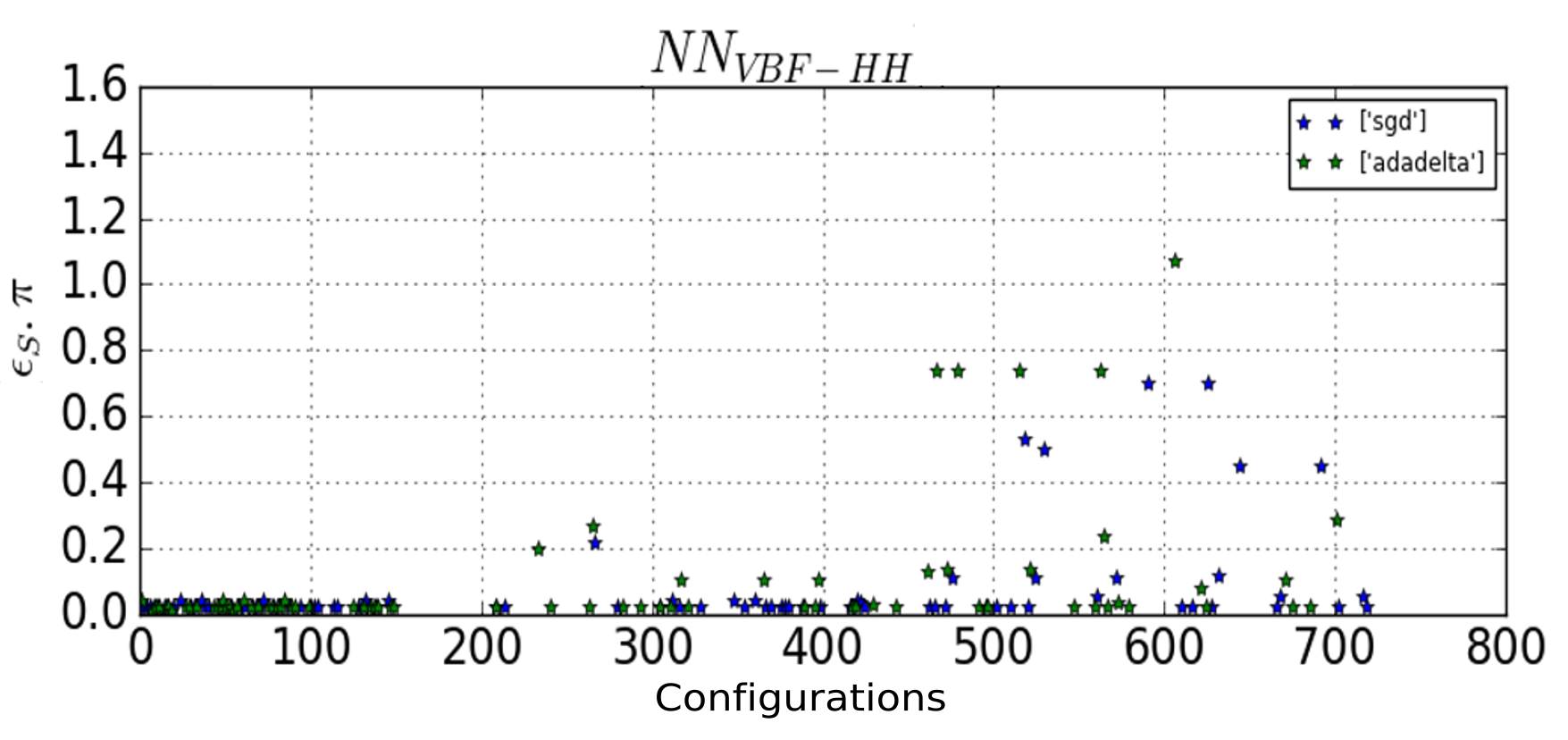}
		\caption{} \label{fig:5b}
	\end{subfigure}\hspace*{\fill}\\   
	\hspace*{\fill}\begin{subfigure}{0.49\textwidth}
		\includegraphics[width=\linewidth]{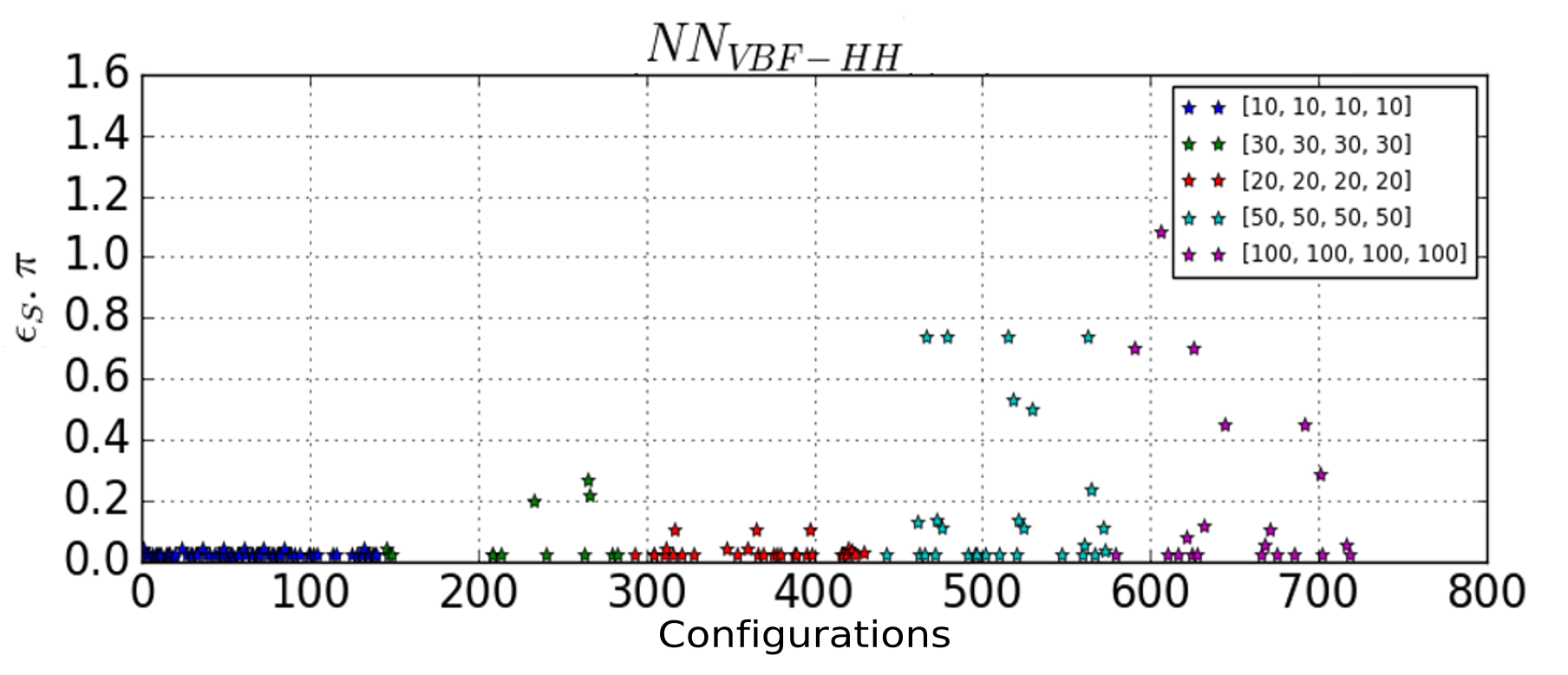}
		\caption{} \label{fig:5c}
	\end{subfigure}\hspace*{\fill}   
	\caption{\small{Examples of Deep Neural Network (DNN) hyper-parameters scan, respectively (\ref{fig:5a}) for the activation function (relu, selu and tanh were tested), (\ref{fig:5b}) minimization loss function algorithm (sgd and adadelta) and (\ref{fig:5c}) the network structure (the maximum number of hidden layers and neurons per each layer are respectively 4 and 1000, while the minimum values are 1 and 10). On the y-axis, the maximum of the purity ($\pi$), which is defined as the ability of the classifier not to label as signal an event that is background, times the signal efficiency metrics ($\epsilon_{S}$), which is the ability of the classifier to find all the signal samples, is plotted for the testing data set. This metrics is shown against the number associated to the i-th trained DNN configuration model (out of 722 DNNs). The best model corresponds to the highest purity times efficiency metrics, i.e. the model 620. }} \label{fig:5}
	\end{center}
\end{figure}
\section{Multivariate Analysis and Results}\label{comments}
Training MVA methods is a procedure that always needs to be done in multiple parallel steps. Indeed, it is hard to guarantee that, for a set of inputs, a set of MVA hyper-parameters are the optimal choice, since there is a huge number of combinations that one can build. Those configurations (including the training data set size) can strongly affect the evolution of the MVA training.

In order to optimize the study, the framework developed in Python has the feature to build up, through Keras, different DNN architectures and perform multiple parallel scans. These can run over different sets of inputs and several parameters that need to be configured for the training (see \autoref{fig:5}).

\begin{figure}
	\hspace*{\fill}\begin{subfigure}{0.32\textwidth}
		\includegraphics[width=\linewidth]{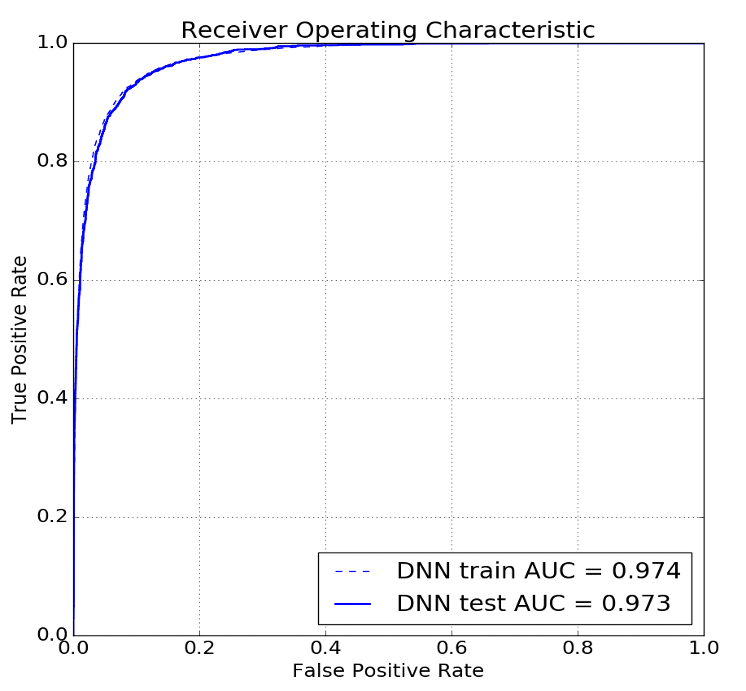}
		\caption{} \label{fig:4a}
	\end{subfigure}\hspace*{\fill}   
	\begin{subfigure}{0.33\textwidth}
		\includegraphics[width=\linewidth]{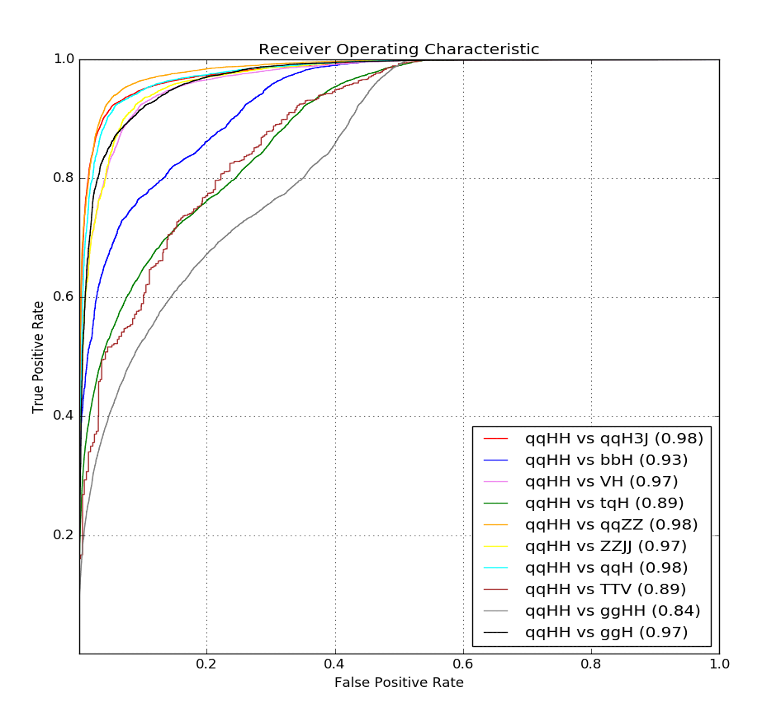}
		\caption{} \label{fig:4b}
	\end{subfigure}\hspace*{\fill}  \\ 
	\begin{subfigure}{0.3\textwidth}
		\includegraphics[width=\linewidth]{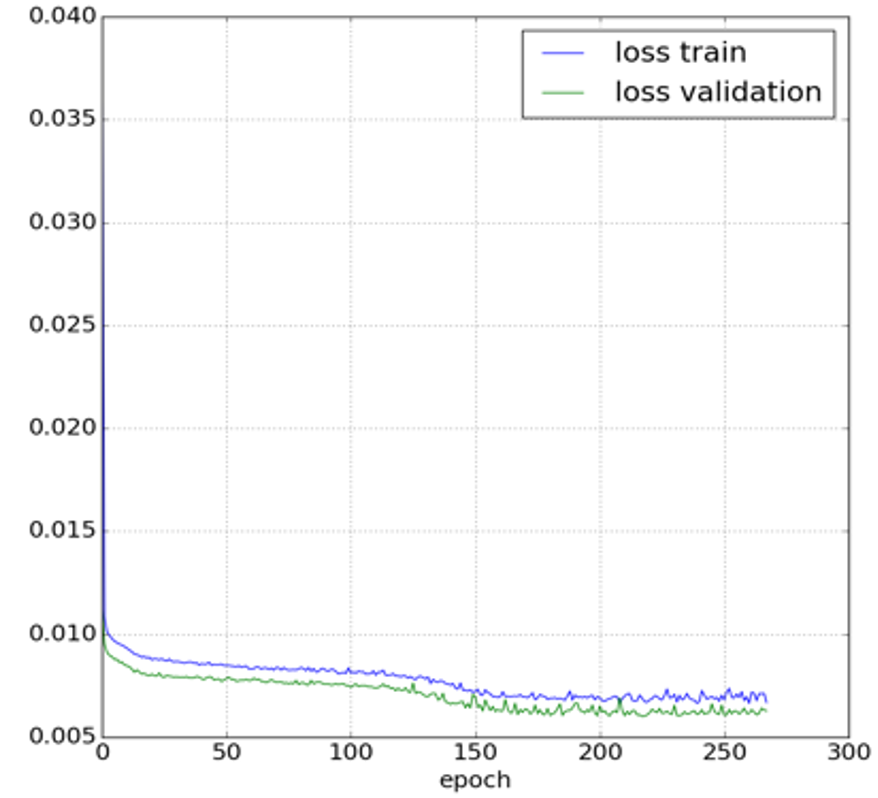}
		\caption{} \label{fig:4c}
	\end{subfigure}\hspace*{\fill}   
	\begin{subfigure}{0.34\textwidth}
		\includegraphics[width=\linewidth]{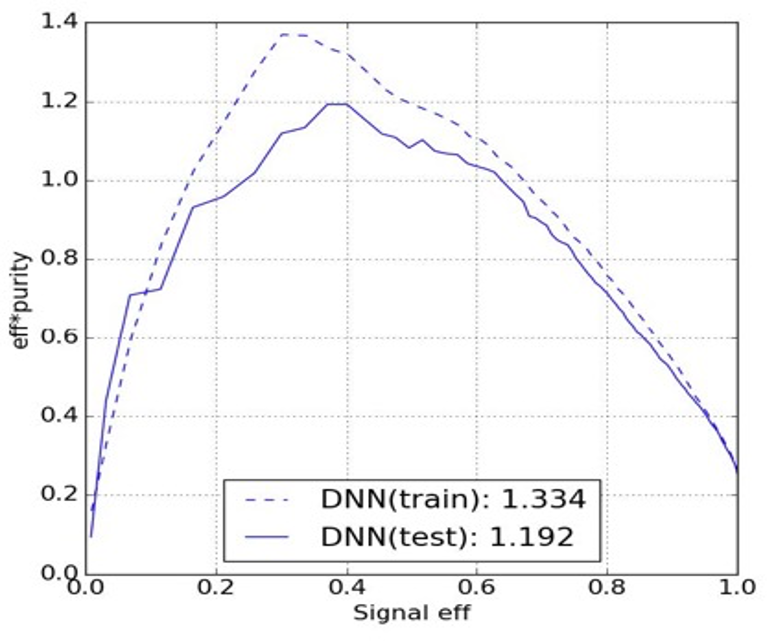}
		\caption{} \label{fig:4d}
	\end{subfigure}\hspace*{\fill} 
	\hspace*{\fill}\begin{subfigure}{0.3\textwidth}
		\includegraphics[width=\linewidth]{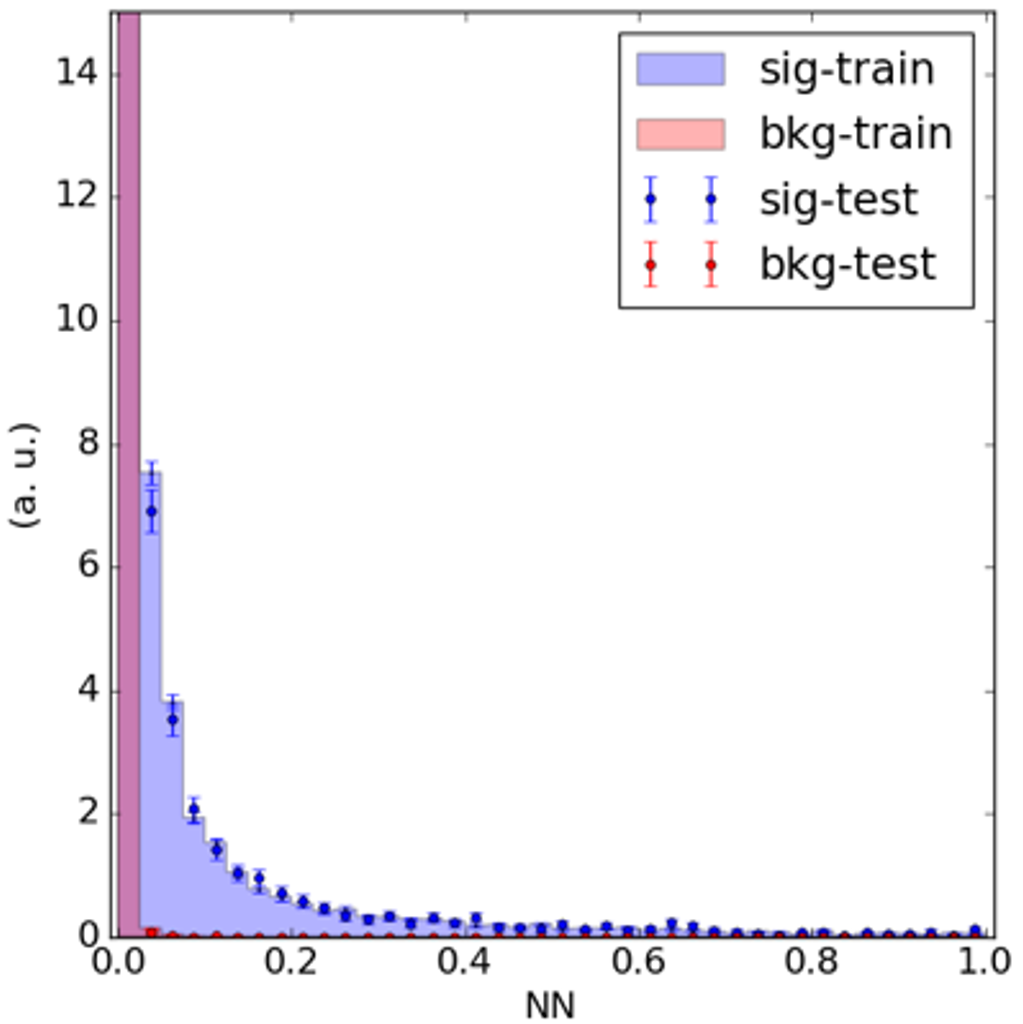}
		\caption{} \label{fig:4e}
	\end{subfigure}\hspace*{\fill}   
	\caption{\small{Discriminating performance plots of the best Deep Neural Network model in terms of the area under the ROC curve (AUC), which is the probability that a classifier will rank a randomly chosen signal instance higher than a randomly chosen background one, for the merged backgrounds (\ref{fig:4a}) and the separated ones (\ref{fig:4b}), the loss function (\ref{fig:4c}), the signal efficiency times purity (\ref{fig:4d}) and the DNN score distributions for the training and the test samples.}} \label{fig:4}
\end{figure}

The result of each NN training can be retrieved to produce plots that are used to validate and classify the quality of each training in terms of several evaluation metrics as it is shown in \autoref{fig:4}. In this work, the maximum of the efficiency times purity metrics, $\epsilon_{S} \times \pi$, is used to choose the best model from a parallel training launched on the ReCaS computing center. In this way, we also made over-training checks by overlapping the test and training curves. 

Despite the signal rarity, good discrimination results were achieved in terms of an area under the ROC curve (AUC) of $ \sim$ 98\%  and an AUC discrepancy of $ \sim$ 0.1\% between test and training data sets (it means no over-fitting issues are present). The reader should note that the associated production of the Higgs boson with a t$\bar{t}$ pair (Htt) is excluded from the training because of the high similarity with our signal. A separate discriminator should be trained for that background. 

\section{Conclusions and Plans}

We presented the optimization and training of a DNN algorithm for discriminating very rare SM Monte Carlo signal events from SM background ones getting optimal results in terms of the most common evaluation metrics used in ML.
A similar binary classification task will be performed  for discriminating the VBF HH production in Beyond Standard Model theories from SM bkgs as we started to do in \cite{thesis}. Since these samples have enhanced cross-sections, and it will be simpler to be selected. A Htt killer should be implemented in the future development of the analysis. 

In addition to the scientific originality of the work itself, from these studies several exercises can be derived for both Analysis and Computing Schools, ensuring a deeper understanding of both the particle physics itself and the use of the most suitable ML tools. These techniques have been already applied on a different analysis by the authors and accepted to Hackathon INFN 2021 Competition, targeting master and Ph.D. students \cite{Hackathon}. Indeed, the organization of educational ML hands-on tutorials using the very latest particle physics analyses for young learners will be extended to the discussed double Higgs boson decay channel.

\ack

Authors would like to thank IT resources made available by ReCaS, a project funded by the MIUR (Italian Ministry for Education, University and Re-search) in the “PON Ricerca e Competitività 2007–2013-Azione I-Interventi di rafforzamento strutturale” PONa3\_00052, Avviso 254/Ric, University of Bari.

\smallskip
\end{document}